# Against racing to AGI: Cooperation, deterrence, and catastrophic risks

Leonard Dung and Max Hellrigel-Holderbaum

**Abstract**: AGI Racing is the view that it is in the self-interest of major actors in AI development, especially powerful nations, to accelerate their frontier AI development to build highly capable AI, especially artificial general intelligence (AGI), before competitors have a chance. We argue against AGI Racing. First, the downsides of racing to AGI are much higher than portrayed by this view. Racing to AGI would substantially increase catastrophic risks from AI, including nuclear instability, and undermine the prospects of technical AI safety research to be effective. Second, the expected benefits of racing may be lower than proponents of AGI Racing hold. In particular, it is questionable whether winning the race enables complete domination over losers. Third, international cooperation and coordination, and perhaps carefully crafted deterrence measures, constitute viable alternatives to racing to AGI which have much smaller risks and promise to deliver most of the benefits that racing to AGI is supposed to provide. Hence, racing to AGI is not in anyone's self-interest as other actions, particularly incentivizing and seeking international cooperation around AI issues, are preferable.

## 1. Introduction

Many researchers and decision-makers believe that generally intelligent artificial agents (AGI) will be developed within the next 10 years. Such systems can autonomously act over long time scales in the pursuit of complex goals. Moreover, by definition, they are in all cognitive capacities (reasoning, planning, etc.) and cognitive domains (mathematics, natural science, literature, engineering, etc.) at least as proficient as the most competent humans. More generally, recent years of AI progress reflect a trend of fast advances in AI capabilities, in line with several theoretical models and predictions of rapid future progress (Barnett 2023; Cotra 2020; Davidson 2021; Grace et al. 2023; Kwa et al. 2025). It hence seems likely that AI systems will continue to advance quickly in specific and general capabilities. On the view of interest here, such AI systems will eventually be sufficiently powerful to provide a decisive strategic advantage (DSA), meaning that whichever actor (state, company, etc.) controls them can use them to take control over others, for instance by overwhelming them militarily. While we focus on AGI, other levels of (general or specific) capabilities may be necessary and/or sufficient for obtaining a DSA, such that similar considerations apply.

These assumptions have motivated a variety of actors to advocate for a "race to AGI" – i.e. for investing significant (and increasing) resources into AI development and deprioritizing concerns about safety to be the first to develop AGI. While many have already pointed to the fact that such race dynamics increase the risks associated with AGI development, influential political actors have increasingly adopted the view that an AGI race is very likely to ensue,[1] and that their nation should try to win it.

In this paper, we argue that racing is in no actor's interests, including the likely winner of the race. A distinctive feature of our argument is that it nonetheless grants many central assumptions that proponents of racing to AGI make. In section 2, we reconstruct the view that some state – typically the US – should race to AGI and analyze the central assumptions motivating it. In section 3, we argue that racing to AGI would substantially increase a multitude of risks (including for the likely winner). Therefore, the expected downsides of racing to AGI are much worse than admitted by proponents of racing. We further argue in section 4 that the chances of realizing the private benefit of racing, namely achieving a DSA, may be lower than it seems, even if one develops AGI first. Combining both points, section 5 argues that racing to AGI is generally in no one's self-interest. While racing may be warranted if one's opponent is certain to race, this assumption is implausible and, furthermore, one can influence whether one's opponent races. In particular, as we contend in section 6, deterrence and promoting various forms of cooperation can influence whether one's opponents will race. As cooperation promises additional substantive upsides, these alternatives should be preferred to racing to AGI. Section 7 concludes.

## 2. AGI Racing
### 2.1 Proponents of AGI Racing

In public, leaders of the most influential AI companies have pronounced that they expect AGI (or some similar form of powerful AI) soon. Dario Amodei – CEO of Anthropic – wrote that AI agents which are smarter than a Nobel Prize winner across most relevant fields could be developed "as early as 2026".[2] Sam Altman – CEO of OpenAI – said in a recent interview that OpenAI is "confident we know how to build AGI as we have traditionally understood it".[3] Demis Hassabis – CEO of Google DeepMind – expressed that human-level AI is "probably three to five years away".[4] Such strong predictions have also been made by others who do not work at for-profit companies (Kokotajlo et al. 2025). While not necessarily making equally bold predictions, substantial concern about the potential development of AGI has also been voiced by leading AI researchers at universities such as Stuart Russell (Russell 2019), Yoshua Bengio, and Geoffrey Hinton (Center for AI Safety 2023).

---

[1] While the rhetoric has shifted in this direction, and at least some seem to believe that a race has started already, we do not think that we are currently in a race to AGI (see section 5).
[2] *https://darioamodei.com/machines-of-loving-grace*.
[3] https://time.com/7205596/sam-altman-superintelligence-agi/
[4] https://www.bigtechnology.com/p/google-deepmind-ceo-demis-hassabis

Most of these industry leaders, as well as some political actors, advocate for racing to AGI. Altman and Amodei both say that a more democratic world depends on democratic countries, like the US, developing AGI before authoritarian states, like China.[5] AI companies, like OpenAI, argue against specific AI regulations – such as regulations concerning the use of copyrighted materials – by saying that it would damage the US' prospects in the AGI race against China.[6] The U.S.-China Economic and Security Review Commission's 2024 report advised Congress to set up a "Manhattan Project-like program dedicated to racing to and acquiring an Artificial General Intelligence (AGI) capability."[7] The US' recent AI Action Plan includes "Winning the Race" in its title while stating "it is imperative that the United States and its allies win this race".[8] An AGI race between US and China is also predicted, not endorsed, by a detailed scenario about AGI development constructed by AI researcher Daniel Kokotajlo and colleagues (Kokotajlo et al. 2025). Finally, Leopold Aschenbrenner, in a prediction of the next decade of AI which has reportedly been influential in US policy circles, captures the essence of the view (Aschenbrenner 2024):

> Superintelligence will give a decisive economic and military advantage. China isn't at all out of the game yet. In the race to AGI, the free world's very survival will be at stake. Can we maintain our preeminence over the authoritarian powers?

## 2.2 Analyzing the AGI Racing view

The view of concern can be expressed as follows:

AGI Racing: A contextually relevant actor (most often: the US government) should "race to AGI". One races to AGI if one's actions relevant to (frontier) AI development are primarily driven by the goal to speed up one's AI development and thereby achieve AGI faster than other actors.

AGI Racing thus motivates – absent other overwhelming opposing reasons – two key priorities: First, one ought to refrain from taking actions which may slow down one's own frontier AI development by imposing rules on it or by voluntarily constraining oneself. Second, one ought to try to slow down one's competitors' frontier AI development.

---

[5] https://time.com/7205596/sam-altman-superintelligence-agi/ and *https://darioamodei.com/machines-of-loving-grace*.
[6] https://openai.com/global-affairs/openai-proposals-for-the-us-ai-action-plan/
[7] https://www.uscc.gov/annual-report/2024-annual-report-congress see also https://www.congress.gov/event/119th-congress/house-event/118428 for similar sentiments.
[8] https://www.whitehouse.gov/wp-content/uploads/2025/07/Americas-AI-Action-Plan.pdf

Note that AGI Racing does not need to be defined in terms of AGI (as we have defined it). Other forms of AI which are powerful enough to have transformative impacts may also suffice, and a race may naturally begin before the development of such systems is imminent and continue subsequently. More precisely, any form of AI which is powerful enough to provide a decisive strategic advantage (DSA) makes a plausible target of an AI race under AGI Racing. A DSA is "a level of technological and other advantages sufficient to enable [one] to achieve complete world domination" (Bostrom 2014, p. 78).

AGI Racing is typically motivated based on four key assumptions. The first assumption is that building AGI (or some other very powerful form of AI) soon is technically feasible. Many would be inclined to place less weight on the importance of AGI development if AGI lies far in the future.

The second assumption is that successfully controlling AGI, while competitors do not, provides a DSA to whoever controls AGI. The general thought here is that AGI enables the automation of most or all labor and that large groups of AGI systems working together can massively speed up economic and technological progress as well as enhance societal decision-making. Massive economic and technological superiority, then, translate into military superiority.[9]

The third assumption is that one's national security and prosperity crucially depends on which actor obtains a DSA through AGI. As we have seen, the most common fear is that an authoritarian state might use this DSA to achieve lasting world domination, thus spreading human-rights abuses, abolishing free and fair elections, and entrenching totalitarian population control.[10] Even if that would not happen, it seems clear that an opposing state with a DSA would be a grave threat to national security.

Notice that AGI Racing (as we defined it) focuses on national self-interest. On this view, nations should race if racing is in their self-interest. Instead, the view that an actor A, *morally*, should race depends, given a cosmopolitan notion of morality, on the assumption that A winning the race is much better for the world. For example, one may hold that it is morally desirable that the US wins an AGI race if one holds that it is substantially less likely that liberal democracies would use their DSA for nefarious purposes. Such states might either refrain from overly interfering with other states or, alternatively, spread and entrench liberal democracy.

---

[9] What if one does not believe that powerful AI may provide a DSA? While the stakes of developing powerful AI would be lower, one could clearly still hold that progress in AI would have substantial benefits, also with respect to international competitors. We, however, do not count this as an instance of AGI Racing for two reasons. First, a race has a clearly defined goal state (e.g., there is no GDP race). A state in which a DSA is achieved constitutes such a goal state while it is not clear which other state could play this role. Second, and relatedly, losing the AGI race loses much of its anticipated horror if AI does not provide a DSA. For, even if one loses, other nations (at least jointly) cannot be dominated by the winner and may catch up in time.

[10] See e.g. https://a16z.com/ai-will-save-the-world/: "China has a vastly different vision for AI than we do – they view it as a mechanism for authoritarian population control, full stop. [...] The single greatest risk of AI is that China wins global AI dominance and we – the United States and the West – do not."

We will be concerned with the self-interest version of AGI Racing because it is less demanding. The same arguments we use below also undermine the view that racing is *morally* appropriate.

The fourth assumption in support of AGI Racing is that some other actor is also racing to AGI or is likely to race in the future. If other actors, e.g., China, are not planning to build AGI rapidly, then there is no urgent need to race to AGI oneself.

For the sake of the argument, we accept assumptions 1-4 in section 3. This way, we show that – even given key ingredients of the worldview of AGI Racing proponents – AGI Racing faces severe challenges. Later, we critically discuss assumption 2, i.e., whether AGI will provide a DSA (section 4), as well as assumption 4 – that other actors are racing to AGI (section 5). The four previous assumptions may suffice as a characterization of what drives AGI Racing. To subsequently address the strongest version of this position however, we now turn to three common arguments in support of AGI Racing.

Spelling out those arguments requires some context: Many proponents of AGI Racing grant that there are substantial risks in developing AGI. Most prominently, it is often believed that wrongly designed AGI may resist human control and seek power, leading to AI takeover, permanent human disempowerment and perhaps human extinction (e.g. Bostrom 2014; Cappelen et al. 2025; Carlsmith 2022; Center for AI Safety 2023; Dung 2024a; Ngo et al. 2023). A popular response, especially promoted by leading AI companies (Friederich & Dung 2025), is that we should first solve the alignment problem and then only build AI systems which are "aligned". Roughly, systems are aligned with their users and/or designers if (and only if) they act as intended by their users and/or designers. The alignment problem can be understood in a variety of ways (Dung 2023; Gabriel 2020; Ngo et al. 2023), but it is often framed as a challenge which can be solved predominantly with the tools of technical computer science (Friederich & Dung 2025). With that, the three supporting arguments go as follows.

First, AGI which is aligned to human goals and is used for good has high welfare benefits. The same reasons why AGI would provide a DSA are also reasons why AGI might substantially increase human welfare. AGI labor could be used to massively increase economic growth, cure diseases, counteract aging, and improve individual and societal decision-making, just to name a few likely benefits (Amodei 2024; see also (for someone who does not promote racing) Russell 2019, ch. 3). If aligned AGI promises very high welfare benefits, then – all else being equal – it is better to receive these benefits earlier than later. Hence, the benefits of AGI provide a reason to speed up frontier AI development.

Second, progress in AI alignment research may be mostly driven by progress in frontier AI. That is, progress in knowing how to align AI depends on more capable AI systems being available. If so, then slowing down AI development might not provide much benefit for AI alignment research – since it would also slow down alignment research. If progress in alignment research is the main lever for designing AI which is overall good for the world, then this would substantially reduce the benefits of slower AI development. This entails that the downsides of an AGI race are reduced.

There are two reasons to think that frontier AI research is central for progress in AI alignment research. For one, such progress provides alignment researchers with more powerful systems to study. In particular, to avoid AI takeover, research needs to figure out how to align AI systems which possess human-level or superhuman capacities in many domains. When studying less capable systems, it is often hard to ascertain whether relevant results generalize to more powerful systems. Another consideration is that highly capable AI systems can themselves make contributions to AI alignment – as researchers, not just as objects of study. OpenAI's alignment plan is (or was)[11] based on building a human-level automated alignment researcher which can then be used to align more powerful systems (Burns et al. 2023; OpenAI 2023). Google DeepMind's "Approach to Technical AI Safety and Security" also emphasizes the potential of future AI to improve human oversight or even to replace the need for human oversight via automated methods (Shah et al. 2025, sect. 6).

A third supporting argument for AGI Racing builds on the assumption that the speed of AI progress may at some point massively accelerate, since AI systems might themselves increasingly contribute to frontier AI development. In a postulated "intelligence explosion" (Chalmers 2010; Bostrom 2014; Davidson 2023, Eth & Davidson 2025, Kokotaljo et al. 2025) advances in AI recursively feed back into AI research which in turn leads to further AI advances and so forth. Expecting an intelligence explosion supports the assumption that it is technically feasible to develop AGI soon and that AGI provides a DSA. Given this view, once AI can substantially accelerate AI research itself, AGI and a DSA are close (for critical discussion, see section 4).

---

[11] OpenAI has disbanded their "Superalignment team" in 2024. It was also reported that OpenAI violated their commitment of using 20% of their secured computational power for the superalignment team: https://fortune.com/2024/05/21/openai-superalignment-20-compute-commitment-never-fulfilled-sutskever-leike-altman-brockman-murati/. Hence, this plan may be obsolete now.

So, overall, the benefits of AGI, the assumption that alignment research progress depends on frontier AI research progress, and the possibility of an intelligence explosion support rapid AI development and AGI Racing. As a means of illustrating the view, consider the following matrix:

| Agent A / Agent B | Agent A does not race (*cooperates*) | Agent A races (*defects*) |
|---|---|---|
| **Agent B does not race** (*cooperates*) | Agent A: 3 utils (N) <br> Agent B: 3 utils (N) | Agent A: 4 utils (UR) <br> Agent B: 0 utils (UN) |
| **Agent B races** (*defects*) | Agent A: 0 utils (UN) <br> Agent B: 4 utils (UR) | Agent A: 2 utils (R) <br> Agent B: 2 utils (R) |

**Table 1:** Classical prisoner's dilemma adapted to the pace of AI development according to AGI Racing, where the variables have the following meanings: N: Not Racing, R: Racing, UR: Unilateral Racing, and UN: Unilateral Not-Racing. As for the prisoner's dilemma: UR > N > R > UN. Hence, as N > R, mutual cooperation is superior to mutual defection. However as UR > N and R > UN, defection is the dominant strategy for both A and B.

This matrix, and the ones we use subsequently, focus on characterizing the ordinal ranking of options implied by different views correctly. One should not place much weight on specific numbers beyond their ordering. Importantly, however, the matrix shows that, on AGI Racing, the strategic situation with respect to the pace of AI development mirrors a prisoner's dilemma. That is, the dominant strategy for both actors is to defect, that is, to race. This is a central tenet of AGI Racing to which we object below.

In the next section, we present three arguments against racing to AGI which focus on the potential downsides of racing. Later, we argue that the upside of racing may be overestimated and, crucially, that racing increases the chance of opponents racing to AGI (section 5). Jointly, these arguments, in conjunction with the alternatives to racing (section 6), suggest that racing is not in one's self-interest, contrary to the previous considerations. We do not claim that our arguments against AGI Racing are exhaustive. Katzke and Futerman (2024) have provided further, and to some extent related, criticisms. A central assumption underlying our arguments in the next section is that racing increases various risks. When racing, speed has high importance, but rapid development threatens the possibility of cautious development and deployment as well as societal adaptation to changes. Racing hence involves a willingness to sacrifice some amount of safety in exchange for competitive advantages. Many authors have argued for this assumption in the past (Armstrong et al. 2016; Hendrycks et al. 2023; Katzke and Futerman 2024; see also Cave & Ó hÉigeartaigh 2018; Emery-Xu et al. 2024).

## 3. The downsides of an AGI race
### 3.1 The many risks argument

For concreteness, let us call a risk "catastrophic" if it is the chance of an event, or a set of related events, which causes the death of at least 100 million human lives, or something which has comparable negative moral significance (Dung 2024b). We hold that there are multiple credible independent catastrophic risks from AI. A central goal of policy should be to avoid catastrophic risks of any kind. To specify these risks, we rely on previous hypotheses from the literature (for related lists, see Critch & Russell (2023), and Hendrycks et al. (2023)):

1. We already talked about *takeover through misaligned AGI*. To rehearse, the idea is roughly that (by default) we may create AGI systems with goals that are at odds with human values. If so, such systems may in service of their goals learn to pursue dangerous subgoals like seeking power or accumulating resources since those are generally useful for most goals. Since such systems are by definition more powerful than us, they may plausibly disempower humanity.

2. *Catastrophic aligned AGI misuse*. If AGI provides a DSA, then a group or an individual which obtains control over AGI could use it to achieve world domination and use this for malevolent ends (Friederich 2023; Katzke and Futerman 2024; Yum 2024).

3. *Preventive war*. If AGI is believed to provide a DSA, then nations have an incentive to wage war to prevent their adversaries from obtaining AGI (Hendrycks et al. 2025; Katzke and Futerman 2024). This is particularly plausible if their adversaries will not, or cannot, credibly pre-commit to not using AGI to achieve domination over them.

4. *Accumulative catastrophic risk*. Rapid advances in AI may cause successive disruptions which lower the stability and response capacity of different societal systems until a cascading breakdown of these systems leads to irrecoverable collapse (Kasirzadeh 2025; Bales 2025).

5. *Gradual aligned AGI takeover*. Even if individual AI systems behave in line with individual human intentions, competitive incentives and coordination failures may drive humans to gradually increase their reliance on AI and thereby to transfer power to AI systems, until societal systems are decoupled from human feedback and control (Kulveit et al. 2025).

Our argument does not require that all of these hypotheses are accepted; only some of them. Moreover, these five hypotheses rest on substantively different assumptions, so that rejecting one of them does not provide much reason to reject others.[12]

It is also important that these risks are not one-shot but could be realized at different stages of AI development, including stages long after AGI has been developed. It may, for instance, turn out that the first AGI is not misaligned to an extent leading to catastrophic consequences, but that such misalignment arises later, with even more powerful systems. Similarly, catastrophic AGI misuse might be successfully prevented for some time, but then happen later anyway – when AI systems or related technologies, institutional contexts, or relevant actors change. So, the same type of risk – for instance, catastrophic AGI misuse – is diverse in the sense that it could be realized at very different stages of AI development.

*Why, then, does the assumption that there are multiple independent catastrophic risks of AGI development speak against AGI Racing?*

Most obviously, if there are multiple independent risks from AI development which could lead to a catastrophe, this raises the probability that AGI development will actually cause a catastrophe. Per our previous assumption, since racing involves deprioritizing risk-mitigation relative to the goal of building AGI first, catastrophic risks are higher than otherwise, so that there is a strong reason against racing. Crucially, while most of the risks are already present when one's adversary is racing, not oneself, one's own actions influence how likely the adversary is to race (see section 5 and the subsequent discussion of deterrence and cooperation).

More specifically, we have seen that AGI Racing is supported by the assumption that progress in technical alignment research can make AI systems safe, and thus prevent catastrophic outcomes. However, achieving a technical understanding of AI alignment and implementing it in all frontier AI systems by itself only eliminates the first risk. While being able to rely on aligned AGI may help address the other four risks, they are not premised on the idea that AGI is misaligned. Thus, AI alignment alone is insufficient to address them.

The observation that these risks are not taken care of once one has aligned AGI also challenges the characterization of AGI development as a winner-takes-all race. The idea that AGI provides a DSA is most naturally expressed as the view that, once one successfully controls AGI, one has the power to safely use it to achieve whatever political, economic, military, etc.

---

[12] One may complain that these hypotheses, for example, are all motivated by the shared assumption that substantial increases in AI capacity are likely. However, this assumption is crucial to AGI Racing, which is why, for the sake of the argument, we presuppose it here. The risks may even be anti-correlated in the sense that, for example, learning that takeover through misaligned AGI is not realistic may increase the probability of catastrophic AGI misuse (Hellrigel-Holderbaum & Dung, forthcoming).

purposes one intends. However, if – even at that point – risks of catastrophe loom, this restricts how AGI can be wielded, provided one wants to avoid catastrophic outcomes, including those affecting oneself. We now briefly consider one catastrophic risk from frontier AI development in more detail.

### 3.1.1 Nuclear instability

Let's consider nuclear instability as an underrepresented catastrophic AI risk. Here is the core concern: AI development could undermine the classical mutually assured destruction (MAD) regime and thereby the current equilibrium in nuclear security and deterrence. An AGI race further exacerbates this threat (more on that below). We see at least three pathways for AI development to undermine MAD and similar forms of nuclear deterrence in addition to the aforementioned preventive war risk:

First, AI development will lead to increasingly fast, strong, numerous, and widespread autonomous weapons systems,[13] e.g. small AI-steered drones. This may enable militaries (and potentially private actors) to cause havoc in a host of ways without obviously having crossed a red line regarding nuclear warfare. Eventually, adversaries may be drawn to (threatening) using military force to stop such attacks, hence substantially increasing the risk of nuclear escalation. In addition, autonomous weapons systems could attack and potentially knock out command and control capacities, as well as nuclear weapons themselves especially quickly, hence endangering second-strike capabilities characteristic of MAD (Horowitz 2019).

Second, AI-based intelligence operations may make it much easier to determine the positions of nuclear warheads, intercontinental ballistic missiles, and nuclear submarines. If an actor knows the position of their opponent's nuclear weapons, then it becomes increasingly feasible to engage in preventive strikes with a much lower risk of nuclear retaliation. This, or possibly the mere impression that adversaries have such capabilities (Geist & Lohn 2018), would undermine MAD. As it becomes more feasible to preventively knock out nuclear weapons, all nuclear powers need to assume that due to this, the risk of losing their nuclear capacity increases. While hence inching towards hair-trigger alert, nuclear powers have to take increasingly ambiguous signs of preventive strikes seriously.

Third and finally, AI-based or AI-facilitated decision-making in militaries could raise the chances of major wars, possibly including nuclear warfare with countervalue targeting.[14] Rivera et al. (2024), e.g., recently presented empirical evidence that current LLMs have quite worrisome tendencies regarding nuclear escalation, while Geist & Lohn (2018) consider that AIs' vulnerability to adversaries' influence could result in the subversion of decision-making involving AI systems; see also Maas et al. (2023) for more general considerations. A particularly concerning trend here involves decreasing times for military decision-making which may increase the risk of errors or failures of judgement and undermine de-escalatory means such as calls over direct communication links such as "hotlines".[15]

---

[13] See e.g. https://autonomousweaponswatch.org/weapons for some examples of current systems.
[14] For a recent illustration in a short movie, see https://www.youtube.com/watch?v=w9npWiTOHX0.

Now, why is this particularly dangerous given an AGI race? Most generally, an AGI race leads to a deprioritization of risk-mitigation efforts and reduces the time available to limit nuclear risk in particular by, e.g., de-escalating crises, or negotiating and reaching agreements. How can general AI systems influence the relevant military capacities? They could either be put to military uses (e.g. for intelligence operations) or contribute to progress in narrow military AI, e.g. by accelerating scientific progress generally. Finally, a forceful competitive dynamic including a potential arms-race e.g. in autonomous weapons, increases tensions between opponents as well as the chance of escalation while decreasing the chance of international cooperation.

Overall then, the plurality of risks from hastened AI development – including nuclear instability – provides a strong reason against racing to AGI given the potential downsides.

## 3.2 The social risks and capacities argument

Many of the five catastrophic risks mentioned above resist a fully technological solution. Instead, addressing them depends, at least in part, on social responses such as: solving coordination problems, constraining the influence of dangerous actors, easing political tensions, changing citizens' values, or increasing the resilience of social systems. As we have seen, one supporting argument for AGI Racing is that progress in frontier AI development is crucial for progress in technical alignment research. If this were so, it would imply fewer benefits to developing AGI slowly, and avoiding racing, than otherwise.

However, even if this is the case, the view that risk mitigation strategies depend crucially on progress in frontier AI research is less plausible once we consider risk mitigation that depends on social responses, not just technical ones. We concede that frontier AI systems can play some role here too: For example, there is evidence that current LLMs are able to facilitate and improve political deliberation (Tessler et al. 2024).[16] Yet, it is implausible that technical progress in machine learning is sufficient for solving the following issues: 1. Preventing bad actors from getting access to AGI systems to prevent catastrophic misuse. 2. Achieving a political agreement between China and the US which disincentivizes the loser of an AGI race to start a preventive war. 3. Preventing competitive incentives from leading to the gradual, and individually voluntary, transfer of power from humans to AI systems. These are only a few of the social challenges implicated in mitigating the catastrophic risks mentioned above.

---

[15] On the other hand, AI-based or AI facilitated decision-making could also reduce such risks, e.g. because the relevant systems may be (trained to be) particularly reliable or rational in their decisions/recommendations.

[16] Though conversely, they may also undermine epistemic and democratic processes; see e.g. Bales (2025) and Seger et al. (2020).

Instead of being supported by fast AI development, many social processes are slow and cannot be sped up arbitrarily. For example, especially in democratic societies, enacting legislation takes time. Ideally, such legislation should build upon an extensive deliberative process, among political decision-makers as well as society at large (Bächtiger et al. 2018). Cultural processes and institutions also take time to adjust to changes, for instance changes due to technology. Another issue is that an AGI race incentivizes secrecy to prevent competitors from copying important research ideas, which conflicts with the transparency required for public deliberation and accountability to the public. Hence, to the extent that social processes are crucial for addressing catastrophic risks from AGI development, racing to AGI increases risk, because it threatens to undermine social responses and adaptation, as they require time and transparency.

### 3.3 The safety research dependence argument

According to AGI Racing progress in technical alignment and safety research is the key to avoiding catastrophic outcomes due to misaligned AI. Yet, we will give two arguments that the conditions of success for many such research avenues presuppose substantial capacities for stopping, inhibiting, and steering AI development and deployment; call these henceforth, following Carlsmith (2025), "capability restraint". Capability restraint is weakened by an AGI race due to its competitive pressures. To spell out the first argument, consider Carlsmith's (2025) toy model of AI safety. In this model, AI is deployed safely if (and only if) the *safety range*, that is the capability level an actor can develop or deploy safely, is above the *capability frontier*, i.e., the capability level this actor has developed or deployed so far. The safety range is advanced by progress in technical AI safety research and the capability frontier is advanced by progress in overall frontier AI research. Then, *risk evaluation* serves to track and forecast the safety range and the capability frontier.

In this model, *capability restraint* plays a crucial role. Whenever risk evaluation indicates that the capability frontier is crossing or will cross the safety range, capability restraint is necessary to avoid unsafe development or deployment. If capability restraint is not possible, then risk evaluation loses much of its value. There is little to gain by knowing that certain AI development or deployment is unsafe if one is unable to stop the development or release of the resulting system in any case.

Much of AI safety research is devoted to risk evaluation (Anwar et al. 2024, Grey & Segerie 2025). This includes evaluations of dangerous capabilities (Balesni et al. 2024; Benton et al. 2024; Shevlane et al. 2023; Weidinger et al. 2023), such as deception (Carranza et al. 2023; Greenblatt et al. 2024; Hagendorff 2024), forecasting of future AI progress (Barnett 2023; Cotra 2020; Wynroe 2023), and alignment audits: tests of whether models have misaligned objectives (Marks et al., 2025). In addition, some methods which might not be classified as risk evaluation depend on capabilities remaining in a certain range, and hence, eventually, on capability restraint. For example, AI control methods may offer safer deployment (Bhatt et al. 2025; Greenblatt et al. 2024) but as most researchers admit, are limited to a certain range of capabilities (Korbak et al. 2025). In addition, methods for achieving effective oversight over future AI systems – roughly, knowing how AI systems behave and understanding the factors driving their behavior (Shah et al. 2025) – depend on the possibility of capability restraint if the oversight methods detect, for example, that a system acts maliciously.

Hence, all these research avenues are much less useful when capability restraint is not feasible. Large parts of technical AI safety research – heralded by AGI race proponents as the means to make the race safe – are therefore themselves diminished when an actor races.

There is another, more general way in which capability restraint supports safety research. Faster AI development and deployment decreases – all else being equal – the chance of solving the core technical challenges necessary to reduce catastrophic risks from misaligned AI. The reason is straightforward: Assume for simplicity that when an actor races, AI development at any point in time is twice as fast as it would otherwise be. Then AI systems at any given level of capability can only be used half as long to advance technical safety research, as they will be replaced twice as quickly. Hence, to generalize: Due to hastened development when actors race, there is a shorter period for AI systems at any present capability level to make progress in technical safety research. Consequently, the expected amount of safety progress is lower at every single capability level, and thus the expected amount of safety progress until each capability level is reached is lower overall. If catastrophic risks typically increase when one reaches higher capability levels, as e.g. on Carlsmith's aforementioned model, then this makes development less safe.[17]

---

[17] See also the distinction between state and step (or transition) risk (Bostrom 2014, pp. 233-234).

Additionally, fierce competition provides incentives to focus most human and AI labor on further increasing capabilities as well as directly fending off adversaries, rather than safety research, further decreasing the amount of labor for AI safety research. AGI Racing is hence wagering on the feasibility of limiting risks from misaligned AI using a reduced total amount of AI labor. Yet, we currently do not know which amount of labor suffices to "solve" the most important technical challenges.[18] It is therefore in one's interest to avoid actions, such as racing, which assume a comparatively lower difficulty of the core technical challenges.[19]

To illustrate the upshot of the arguments presented throughout section 3, we update the matrix introduced above in the situation where both agents race. Note that we remain uncertain whether one should in fact race if one were certain that one's opponent races despite the payoffs suggested below.[20]

| Agent B \ Agent A | **Agent A does not race** (*cooperates*) | **Agent A races** (*defects*) |
|---|---|---|
| **Agent B does not race** (*cooperates*) | Agent A: 3 utils (N) <br> Agent B: 3 utils (N) | Agent A: 4 utils (UR) <br> Agent B: 0 utils (UN) |
| **Agent B races** (*defects*) | Agent A: 0 utils (UN) <br> Agent B: 4 utils (UR) | Agent A: **1 util (R)** <br> Agent B: **1 util (R)** |

**Table 2:** Updated decision matrix regarding an AI race. Due to the foregoing arguments for very severe risks, we reduce the expected value of the outcomes under the scenario in which both agents race (in bold) from 2 to 1.

---

[18] The technical difficulty of eliminating or limiting risks from misaligned AGI is an open and hotly-disputed issue. One reason is that there can be no decisive empirical evidence either way before sufficiently capable AI systems are developed. More concretely, most current alignment techniques, such as RLHF, are expected to work only for a limited range of capabilities (Casper et al. 2023; Dung 2023). See Anwar et al. (2024) and Ji et al. (2024) for overviews of these lines of research. Many commentators see progress on technical safety and alignment research as substantially slower than AI capability advances. If so, extrapolating current trends does not inspire confidence that these issues can be solved with a lower labor capacity.

[19] A further independent argument against AGI Racing is that, even if (other) catastrophic AI risks are avoided, recklessly developing AGI may lock in certain values and thus create a permanently suboptimal state. This is what Will MacAskill calls a "mistopia". https://80000hours.org/podcast/episodes/will-macaskill-century-in-a-decade-navigating-intelligence-explosion/

[20] One consideration, which we bracket henceforth, is that the risks discussed in section 3, such as human extinction, may be (substantially) worse than one's adversary obtaining a DSA. This consideration would suggest that the payoffs stand in a stronger contrast to AGI Racing than captured above.

## 4. The (putative) benefits of racing: Will there be a decisive strategic advantage?

In this section, we analyze an assumption we have granted thus far. There are some reasons to think the chance of realizing the upside that motivates proponents of AGI Racing – achieving a sufficiently large lead ahead of competitors to institute a decisive strategic advantage – is slimmer than proponents of this view generally believe. The reason is this. Imagine current frontier AI is at a certain capability level (CL) 1. It will give rise to CL-2, then CL-3, and so forth. Proponents of AGI Racing seem to imagine that there is some CL-N where AI provides a DSA over anyone with weaker AI, including AI from the previous capability level, i.e. CL-(N-1). However, this is not obvious. AI development may be more continuous in the sense that no single step in capability level provides a DSA. In particular, it is plausible that very advanced AI provides a DSA over competitors with current AI, but such advanced AI may not provide a DSA if competitors have AI that is only slightly behind leading systems. The DSA may instead depend on being sufficiently far ahead of competitors, e.g. having CL-30 while one's competitors are at CL-15.

For a potential example of such a view, Ding & Dafoe (2023) – based on the history of previous general-purpose technologies – see it as rather unlikely that AI advances can or will be turned quickly into military dominance. On this view, while AI may lead to some important military gains, military strength mainly depends on having large numbers of the most important weapons systems such as drones. In light of comparatively slow progress in robotics, one may – at least in the short-term – be skeptical that AI will enable sufficiently rapid manufacturing of relevant weapons and other military technology to attain a DSA.[21] From this perspective, one may further think that a swift and very substantial lead in AI capabilities, sufficient to undermine adversaries' nuclear deterrence, e.g. by granting access to new kinds of weapons or much better (strategic) decision-making, is necessary to gain a DSA. The central question is then whether it is likely that any actor may establish such a substantive lead.

This is an open question. As a skeptical baseline, Brundage & Werner (2025) – for example – point to diminishing returns to scaling up computing power and data to advance AI capabilities. This makes gaining large advantages difficult since it entails that pushing the frontier forward is more difficult than catching up to the frontier, while to catch up, one can additionally make use of competitors' models by e.g. distilling them via an API (see also footnote 22). Nonetheless, future AI systems may enable gains in algorithmic efficiency for training frontier models. Such progress may be independent of (putative) diminishing returns to data and computing power. If so, the next, more advanced AI system could create even more efficient algorithms (and so on) so that fast progress and hence substantial leads could follow, particularly if these jumps in capability are not themselves (soon) diminishing with each step of progress; another key uncertainty. See also Thorstad (2024) for a skeptical analysis and Kirk-Giannini & Davidson (2025) for a rejoinder.

---

[21] It may be particularly important here whether an intelligence explosion is driven by a software or hardware feedback loop; see Davidson et al. (2025a).

We have thus far suggested that establishing a DSA may require a substantial lead over competitors and that establishing such a lead may be hard in practice. We see two further important reasons to believe that getting sufficiently far ahead of competitors to gain a DSA may be hard, which depend on various actions that are open to one's competitors even when falling behind:

First, with current levels of cybersecurity, model theft by state actors cannot be prevented (Nevo et al. 2024; Harris & Harris 2025) and major AI labs will likely not achieve sufficient cybersecurity to rule out this possibility soon. A recent, influential framework differentiates five cybersecurity levels, where the highest level means that one can effectively defend one's operation against top-priority operations by well-resourced state actors. On this scale, AI companies are generally believed to achieve levels 2-3, while reaching 4 or 5 is a substantial challenge that is very unlikely to be achieved by any quick fixes that one could deploy on the order of a few months. If model theft by very well-resourced actors cannot be prevented, then gaining a huge advantage in terms of capability level is unlikely: Competitors may always catch up simply by stealing one's model weights (provided they are also sufficiently skilled at using and further improving the model).[22]

Second, faced with the prospect of being left behind a competitor on AI, nations may – due to the power conferred by advanced AI – make use of sabotage, if not military attacks such as targeted ballistic strikes. If the prospect of one's opponent achieving a DSA is imminent, there may be very strong incentives for such attacks. Hence, in an AGI race, we should expect both sabotage and kinetic attacks to be means that state actors consider to avert AI-enforced subservience by their opponents. Neither, but particularly the latter when coming from a well-resourced adversary, can currently be averted and it seems plausible that either would halt AI developments, say in a particular datacenter, near instantaneously. Hence, sabotage, kinetic attacks, or the threat thereof, can prevent advanced actors from rushing forward with AI development and thus make it harder to achieve a big capability advantage.

---

[22] There are less ambitious operations than stealing model weights which may nonetheless result in a substantial portion of an actor's lead being lost. For example, there have recently been accusations of models being distilled into a competitor's AI system via regular API access; see e.g. https://www.ft.com/content/a0dfedd1-5255-4fa9-8ccc-1fe01de87ea6. Further, the expertise needed to steal model weights may change as AIs themselves can provide relevant expertise, including e.g. for hosting the stolen or distilled model.

None of these considerations are conclusive. For example, it is possible that AI developers increase their cybersecurity standards, helped by AI, to reliably prevent cyberattacks. How likely that is to succeed depends on the (yet unknown) offense-defense balance of AI in cybersecurity.[23] Further, it is uncertain which means actors would be willing to use to thwart the AI development of opponents. Also, even in this case, there is a chance that countermeasures become more sophisticated – though more sophisticated means could also be thwarted by increasingly advanced defenses of AI companies and data centers.

There may be a general objection at this point. Namely, an intelligence explosion – where AI on one capability level rapidly enables reaching the next capability level and so forth – could quickly lead to a big difference in capability levels, hence enabling one competitor to reach a DSA. Yet, even if one grants for the sake of the argument that an intelligence explosion is likely, it is unclear whether it would enable a leading actor to establish a DSA.

The main reason is that a DSA requires that one (collective) agent remains in control of the leading, sufficiently powerful, AI systems. An intelligence explosion, however, increases loss-of-control risks. In particular, risks of AI takeover, accumulative catastrophic risk, and gradual disempowerment – risks not only of catastrophic consequences but risks in which no (collective) human agent maintains a DSA – increase: If these are serious risks and they increase under an AI race, they should likewise increase under a feedback loop of better AIs rapidly (co-)creating better AIs. The reason is simply that as discussed above, these risks increase when AI development happens very quickly and less cautiously. Loss-of-control risk thus provides an additional reason against one nation obtaining a DSA, even if we assume that an intelligence explosion is likely.

Overall, the considerations here suggest that a DSA may be less likely than often portrayed on AGI Racing and is certainly not inevitable. Hence, we should modestly lower the expected benefits of racing when opponents do not race (where the same would hold if opponents would likely lose the race). To (however crudely) summarize these points, we update the previous matrix:

---

[23] In a competition between powerful actors such as nation states, this offense-defense balance seems especially important, as differences in available resources between defenders and attackers are relatively small.

| Agent A \ Agent B | Agent A does not race (*cooperates*) | Agent A races (*defects*) |
|---|---|---|
| **Agent B does not race** (*cooperates*) | Agent A: 3 utils (N)<br>Agent B: 3 utils (N) | Agent A: **3 utils (UR)**<br>Agent B: 0 utils (UN) |
| **Agent B races** (*defects*) | Agent A: 0 utils (UN)<br>Agent B: **3 utils (UR)** | Agent A: 1 util (R)<br>Agent B: 1 util (R) |

**Table 3**: Updated decision matrix regarding an AI race. Based on the arguments casting doubt on the prospect of achieving a DSA, we reduce the expected value of UR from 4 to 3. Thus, N = UR > R > UN. Hence, as N > R, mutual cooperation is superior to mutual defection. Additionally, as N = UR racing is no longer the dominant strategy.

## 5. The comparative danger of racing

We have argued in section 3, using three separate arguments, that the expected downsides of racing to AGI are dire, and much worse than commonly appreciated by proponents of AGI Racing. In section 4, we further argued that the expected upsides of racing to AGI are smaller than commonly portrayed. The upshot of this is that the justification for racing in terms of national self-interest is much weaker than commonly thought. In our view, hence, racing to AGI is generally not worth it, even from an entirely self-interested perspective.

Consider, however, the following objection: One may hold that whether one should race depends on what one's opponents are doing. On this view, as we have argued, if one's opponents refrain from racing, one should do the same. For, by doing so, one avoids the risk from racing while not incurring the risk that an opponent obtains a DSA. However, if one's opponent is racing, one might argue that most of the risks of an AGI race are realized independently of one's actions. Hence, racing oneself only increases the race risks slightly while it substantially decreases the chance of one's opponent obtaining a DSA.

We do not take a stand on whether one should race, given that one is *certain* that one's opponent races, *no matter what happens*. The reason is that these conditions are not realistic. No actor is guaranteed to race. Moreover, if one refrains from racing, one's opponents are substantially less likely to race. There are two reasons for this. First, by racing oneself one changes the opponent's (perceived) incentives. After all, one may convince opponents that racing is necessary to prevent their opponents from gaining a DSA. Relatedly, one increases the chance that one's opponent concludes, falsely, that mutually refraining from racing is not feasible. Moreover, while we are uncertain about this – see footnote 20 for a core source of uncertainty – our matrix as sketched above would imply that the actual incentives, not just the perceived incentives, favor racing only if one's opponent also races. Notice, however, that we are more confident that racing may change perceived incentives for racing, given what current actors believe, which notably is not captured by our matrix.

A central related consideration is that we are not currently in a race to AGI. For instance, there is relatively little investment by governments in advancing cutting edge AI systems, talent still flows between leading labs and countries, and many frontier models are open-weight, all of which would be quite unlikely in an actual race. Ó hÉigeartaigh (2025) further finds that "neither public rhetoric, nor publicly available evidence relating to investment or concentration of relevant resources" support the view that China is currently pursuing an AGI race.[24] If we are not currently in a race, one should be skeptical of the view that it is *inevitable* that one's opponent will race. Instead, there is a risk that one causes one's opponent to start racing, potentially making AGI Racing a self-fulfilling prophecy.

Second, crucially, if one is not racing but one's opponent is, or is advocating for it, there are additional actions available to stop one's opponent from racing. We discuss deterrence and various forms of cooperation which promise additional benefits as two options for making one's opponent less likely to race in the next section.

## 6. Alternatives to racing

In the preceding sections, we argued against racing to AGI. What should be done instead? Table 3 (and table 4 below) reflect that the scenario in which neither agent races is overall preferable and that not racing oneself is a viable strategic choice. However, how can we ensure that opponents do not race, leading to worse overall outcomes? We discuss deterrence and specific forms of cooperation as promising strategies. They also address the concern that opponents may achieve a DSA and hence the purported primary motivation for racing. Let's start with the former.

## 6.1 Deterrence

Recently, Hendrycks et al. (2025) have argued for a position in contrast to AGI Racing: It is in the self-interest of major actors, they claim, to instantiate a deterrence regime called MAIM, short for mutually assured AI malfunction, rather than to race ahead in AI development.[25] The incentives for instantiating MAIM consist in risks from advanced AI and the possibility of one's opponent achieving a DSA, both of which we discussed above. Under MAIM, no one builds superintelligent AI since AI projects aimed at such a goal would likely be 'maimed', i.e derailed using covert sabotage, overt cyberattacks, or kinetic attacks, which would threaten (further) military escalation. Hendrycks et al. hold that using or threatening the use of the just-mentioned actions is much less costly and has a lower chance of failure compared to other ways of reducing the risks posed to states' survival by advances in AI, including massive investments into an AGI race.[26] If all relevant actors are in this strategic position and realize it, this would lead to the implementation of their envisioned deterrence regime.

---

[24] He additionally argues that the race narrative is harmful even beyond potentially causing one's opponent to race as it creates international mistrust which undermines essential opportunities for cooperation on AI. See also our section 6.

We note that AI capability advances are partly driven by algorithmic and hardware improvements. These do not depend on accumulations of big compute clusters in specific locations. Hence, it is hard to see how MAIM would stop these advances. For this reason, it is implausible that deterrence permanently halts rather than decelerates progress in AI, even if it could hinder progress on the most advanced systems for a while. In particular, when enough time passes for gains of increased algorithmic efficiency and hardware to accumulate, including progress in distributed computing to train AIs, big data centers will at some point not be necessary anymore to drive AI development beyond the current frontier.

In the light of the foregoing, deterrence is a path of action deserving further scrutiny, particularly as it may – even if temporarily – prevent progress beyond the cutting edge at whatever point it gets instantiated. It may hence, at least for a while, reduce some of the risks outlined in section 3, so that better social and technical mitigations for these risks can be developed and put in place. Deterrence seems especially important to consider if substantive evidence were to support the imminence of either an AI catastrophe as sketched in section 3 or an adversary reaching a DSA while we (still) have little ability to limit these risks. Further, and due to the incentives under a deterrence regime, deterrence would prevent an AI race and its associated downsides. Nonetheless, we have to admit two major sources of uncertainty which in most circumstances make it unclear whether deterrence should be pursued.

First, deterrence threatens military escalation. Hendrycks et al. are careful to present kinetic attacks as a deterrence option of last resort besides covert sabotage and overt cyberattacks. Yet, whether this option remains a mere last resort is a question that depends on several unsettled empirical questions. In particular, how vulnerable are large-scale and well-resourced, potentially state-backed frontier AI projects to cyberattacks? The security standards for such projects will plausibly be much higher (and rise generally). Further, such projects may be able to employ so many AIs to (largely) autonomously fix bugs that they become rather resistant even to well-resourced top priority projects by state actors. Likewise, physical security may be much higher and increase quickly as the security of such projects is prioritized under higher stakes; see e.g. Harris & Harris (2025). Together with, e.g. more in-depth monitoring enabled by AI-driven progress in surveillance technologies, this may make covert sabotage increasingly infeasible. If frontier AI projects became very resistant to cyberattacks as well as to other forms of sabotage,[27] deterrence in the spirit of MAIM would depend solely on threatening ballistic attacks, resulting in a dangerously unstable situation. It is, after all, easy to imagine how a situation like this could escalate. For instance, one actor may mistakenly believe that their opponent is in a particular data center training the most advanced AI system thus far, perhaps because this opponent has started hiding efforts at advancing frontier AIs. In this case, to uphold deterrence, they may consider firing missiles on said data center, and thereby risk nuclear escalation.

---

[25] We do not count MAIM as an instance of AGI Racing. While slowing down one's adversary is one of the central goals in a race, mutual deterrence results in reciprocal slow down, rather than an acceleration of, one's own AI development.

[26] By taking existential risks from advanced AI and the possibility of one's opponent achieving a DSA to be the central motivation for state action, MAIM may be said to assume a form of defensive realism where states' first priority concerns their survival and security.

There are further ways for deterrence to threaten nuclear escalation: For MAIM to work, i.e. for core stakeholders to decelerate frontier AI development without escalation, these stakeholders have to share some important foundational beliefs. In particular, they have to believe that there is a substantial risk of their project being 'maimed' or of military escalation if they move ahead in their AI development. This belief may mostly depend on believing that one's adversaries hold that frontier AI development beyond a certain point is either extremely dangerous or promises a DSA. In brief: key stakeholders have to already believe in MAIM for MAIM to work. Much as in nuclear security, a central danger in an AI deterrence regime is hence that, there is (always) someone who doesn't get the word.[28] Additionally, frontier AI systems and capabilities are, in contrast to nuclear weapons – the motivating analogy for MAIM

---

[27] It may suffice if only parts of these scenarios manifest for deterrence to become unstable. First, if for *some* relevant adversaries, cyberattacks and sabotage are not feasible, such states may see few viable alternatives to ensure their survival in the face of an opponent threatening to reach a DSA. Second, the same holds if adversaries may be unable to cause *sufficient* damage via cyberattacks or sabotage to a particularly concerning project. For cyberattacks, indeed, it presently seems quite difficult even for well-resourced state actors, to cause severe and targeted, rather than smaller and unspecific damage to adversaries via cyberattacks: even in active conflicts or wars involving states leading in cyber capabilities, there have been rather few very impactful cyberattacks (Smeets 2022; see also https://www.cfr.org/cyber-operations/).

[28] See e.g. the following regarding this phrase in nuclear security: https://nsarchive.gwu.edu/briefing-book/cuba-cuban-missile-crisis/2022-10-27/cuban-missile-crisis-60-most-dangerous-day
https://www.nti.org/risky-business/ask-the-experts-the-60th-anniversary-of-the-cuban-missile-crisis/

– a moving target. As mentioned above, progress in AI hardware and algorithmic improvements is likely to continue such that a deterrence regime is likely to be temporary. Insofar as rival states do not reduce their respective threats, eventual escalation seems quite plausible.

One may respond that a carefully measured form of deterrence can minimize these risks. However, absent means to reliably determine what the right measure is in trying to deter one's adversaries, this provides little comfort.

Do these considerations imply that deterrence ought not to be pursued? We think not. Importantly, we also argued above that a race to AGI would increase the risk of nuclear escalation. Hence, while the risk of escalation rises under a deterrence regime that slows AI development, escalation risk likewise rises under an AGI race. We here cannot determine which of the two incurs a higher risk of nuclear escalation. We note, however, that the risk of escalation is a substantive downside of deterrence. For further discussion of limitations of MAIM and reasons that it may be unstable, see Abecassis (2025), Rehman et al. (2025), and Wildeford & Delaney (2025).

A second argument against MAIM may be that, if deterrence of the kind described remains stable for several years, this may raise the chance of advanced AI systems being misaligned (Ying 2025), thus increasing the corresponding catastrophic risks. First, the incentives under a deterrence regime plausibly make the creation of (strongly) misaligned AIs more likely: deterring one's adversaries may (in the future) be best achieved by pivoting leading AIs against one's adversaries and hence a large number of humans (Kuhn 2025). To remain competitive under a deterrence regime, one needs AI systems that strongly favor oneself while policing rivals to e.g. avoid them racing ahead, and pushing back against them using any means that do not threaten escalation. In a situation where AI systems must prevail over a powerful adversary, they have particularly strong incentives for power-seeking, and human designers for promoting power-seeking. If a deterrence regime makes it more likely that frontier AI develops power-seeking tendencies or makes these tendencies more severe, then human disempowerment through misaligned AI becomes more likely (see section 3.1). That being said, nations would train advanced AI systems to favor their own interest anyhow so that the additional extent to which these systems would be misaligned under a deterrence regime remains unclear.[29]

On the flip side, deterrence, as discussed above, promises to increase the time until very advanced AI systems are developed and thereby the available time for solving key technical challenges such as developing better alignment techniques and protocols. Likewise – insofar as it does not undermine cooperation – a deterrence regime may due to the increased amount of time available, facilitate finding better social mitigations (cf. the social risk argument in section 3.2). Much as for the risk of nuclear escalation, it is thus all in all unclear whether deterrence would on net increase the risks from misaligned AI.

---

[29] Future research should also explore how MAIM interacts with misuse risk, e.g. whether MAIM increases power concentration that may lead to a small group using AI to take over (Davidson et al. 2025b).

**6.2 Cooperation**

Let's then move to cooperation. We believe that international cooperation and coordination are an essential part of any strategy to develop AGI without unacceptable catastrophic risks.

To begin, nations may make enforceable international agreements regarding what safe AGI development requires, perhaps even enforcing them globally: If violating these agreements comes with significant downsides exceeding the costs of satisfying the agreements, then safety-conscious actors have a competitive advantage rather than disadvantage. Thus, the incentives for reckless AGI development are weakened. If the risks of creating AGI (at the current stage of technological and social development) are deemed too high, nations could agree to collectively refrain from building AGI. Short of such decisive measures, nations may impose rules to increase the safety of frontier AI development, without placing direct capability limits on AI development. Such rules could concern mandatory safety testing, monitoring of dangerous capabilities, and many other safety-relevant properties; for proposals and starting points, see e.g. Buhl et al. (2024), Clymer et al. (2024), Paskov et al. (2025), Phuong et al. (2024), and Shevlane et al. (2023).

Besides such jointly-agreed constraints on frontier AI development, there are many more kinds of cooperation which are (largely) independent of each other, underappreciated, and promise substantial benefits. Here is a surely incomplete collection. States or key stakeholders may cooperate to: 1) develop technical AI safety techniques, 2) develop means for preventing severe forms of misuse, 3) evaluate and perhaps establish protocols to align AI systems to human preferences/values (concerning e.g. minimal, shared standards regarding the process or the alignment target), 4) research and develop capacities for societal adaptation to advanced AI (Bernardi et al. 2024), 5) establish AI safety/agent infrastructure (Chan et al., 2025), and 6) verify and enforce crucial properties in frontier AI development; we discuss both shortly.

It is outside the scope of this paper to assess the likelihood that the requisite extent of international coordination and cooperation comes to pass and the necessary policy steps to realize both. We will just make four points: First, substantial international cooperation is possible between geopolitical rivals, as the history of the cold war exemplifies (Bucknall et al.

2025a). Second, if what we said before is correct, international cooperation and coordination are essential. More precisely, without them, it is unlikely that humanity can substantially reduce the risk of catastrophic outcomes and safeguard the attainment of the most desirable post-AGI futures. This is because, without international coordination and cooperation – and especially under the 4 assumptions of AGI Racing – avoiding an AGI race may be hard. After all, if different nations cannot agree to cooperate and AGI stands to provide a DSA, racing seems like one of the few options for nations to protect their security interests (deterrence being another, but only temporary and potentially unstable, one). Additionally, satisfactory mitigations of most if not all of the risks sketched above require adoption by several or all relevant actors, acting in a coordinated manner.

Third, verification and enforcement are two forms of coordination which would make many forms of effective cooperation more likely in practice. "Verification" here refers to methods that provide information about whether an actor complies with a set of agreed-upon obligations, as for example specified in a contract; for approaches see Ammann & Hastings-Woodhouse (2025), O'Gara et al. (2025), Scher & Thiergart (2024), and Wasil et al. (2024). Verification is a central means to build confidence in the compliance of international agreements and treaties.[30]

Enforcement, rather than just providing information about other agents (like verification), changes the payoffs for the party defecting, and hence makes defection more costly. It is clear from our discussion thus far that in the context of developing AGI and potentially reaching a DSA, enforcement mechanisms would need to be very strong, and possibly sensitive to rather small deviations from agreements. There is as of now a paucity of concrete proposals for enforcement methods that would likely lead major actors towards more beneficial actions, but a baseline known from other contexts may be (severe) economic sanctions. Especially if one believes that the appeal of defecting, i.e. racing ahead, is higher than we have suggested, both verification and enforcement are central to decreasing the chance that adversaries defect. Importantly, cooperation thus need not be naïve about adversaries' intent and can just like deterrence (and racing) minimize the risk of opponents achieving a DSA, which again, is the core reason to race on AGI Racing. Note that unlike racing and deterrence, cooperation, if using limited and agreed-upon forms of verification and enforcement, is unlikely to increase the chances of military escalation. Indeed, insofar as cooperation increases trust and benefits the involved agents, it may reduce this risk which is a substantial benefit over alternatives.

---

[30] A central impediment to the use of many forms of verification is that they could unwittingly reveal (some) sensitive information to those verifying compliance (Bucknall et al. 2025a, Katzke and Futerman 2024). Hence, minimizing the amount of information revealed without losing (too much) information about compliance is a central research challenge in developing sophisticated verification mechanisms and regimes (Bucknall et al. 2025b, Scher & Thiergart 2024).
While advancements in algorithmic efficiency may enable building highly capable AI in a way which is harder to monitor, thus constituting a challenge for verification regimes, developing frontier AI may continue to require physically large compute clusters which are relatively easy to monitor.

Fourth, if we are correct that racing to AGI has substantial risks even for the winner of the race, then this provides all relevant nations with a strong incentive to avoid racing (see also Katzke & Futerman 2024 for a similar argument). International coordination and cooperation can reduce the risks from advanced AI that we discussed, which are amplified by an AGI race, and simultaneously reduce the risk that one's opponent obtains a DSA. If so, then the overall incentives favor international cooperation. Hence, if the nations which lead in AI come to understand the risks of an AGI race, then cooperation and international regulation are far from unrealistic.

We can once again illustrate the implications of the feasibility and benefits of cooperation in our matrix:

| Agent B \ Agent A | Agent A does not race (*cooperates*) | Agent A races (*defects*) |
|---|---|---|
| **Agent B does not race** (*cooperates*) | Agent A: **5 utils** (N)<br>Agent B: **5 utils** (N) | Agent A: 3 utils (UR)<br>Agent B: 0 utils (UN) |
| **Agent B races** (*defects*) | Agent A: 0 utils (UN)<br>Agent B: 3 utils (UR) | Agent A: 1 util (R)<br>Agent B: 1 util (R) |

**Table 4**: Final decision matrix regarding an AI race. Since both deterrence and cooperation limit the risks we sketched in section 3, while cooperation may in contrast to racing not increase and even reduce the risk of military escalation, we increase the expected value of N from 3 to 5. Since N > UR > R > UN, we now have a classical trust dilemma (or stag hunt) in which cooperation is both preferable and not dominated (hence making it strategically viable).

To close this section, we want to highlight a hopeful possibility. Deterrence and cooperation may complement each other: A deterrence regime may increase the amount of time available and hence the chances of finding and enacting satisfactory solutions to the core technical and social challenges of frontier AI development, while cooperation improves the chances of finding such solutions per unit of time so that it may eventually offer a (relatively) safe way forward. If so, deterrence would increase the value that cooperation may provide and vice versa.

Of course, it deserves reiteration that deterrence comes with substantive dangers and could, if it led to too much hostility, undermine (most) cooperation. We hence do not currently know whether it is advisable as a strategy, even if it holds some promise.

We consequently conclude that international cooperation is a robustly promising approach for safe frontier AI development. Hence, it should be pursued, in various forms, with determination. Whether deterrence should be pursued remains uncertain at this point.

**7. Conclusion**

Our argument against AGI Racing has proceeded as follows. First, we have shown in section 3 that the downsides of racing to AGI are much higher than proponents of AGI Racing hold. Racing would substantially increase catastrophic risks from AI, including nuclear instability, and undermine the prospects of technical AI safety research to be effective. Second, as argued in section 4, the expected benefits of winning the race may be lower than proponents of AGI Racing hold. In particular, it is not obvious that the winner will gain a DSA over the losers. Third, we argued in the previous section that international cooperation and coordination, and perhaps carefully crafted deterrence measures, are viable alternatives to racing with much lower catastrophic risk. Further, the most significant benefit that racing to AGI is supposed to provide, namely limiting the risk of a DSA by adversaries, equally speaks in favor of these alternatives. Lastly, in limiting the risk of a DSA, cooperation need not raise the chance of military escalation, and may provide independent means for lowering this risk. Hence, given its increased costs and reduced benefits, racing to AGI is not in anyone's self-interest as other actions, particularly various forms of cooperation, are preferable.

**Acknowledgements**: For many helpful comments on earlier drafts of this paper, we thank Alice Helliwell, Ibifuro Jaja, Peter Kuhn, Björn Lundgren, Adriano Mannino, Chris Pang, Caleb Parikh, Matthew Rendall, Ian Robertson, Valen Tagliabue, and Edward Young.

**Author contributions**: Leonard Dung developed the initial idea for the paper. From then onward, Leonard Dung and Max Hellrigel-Holderbaum contributed equally to fleshing out the key idea and to the writing process.

29